\documentstyle[preprint,aps]{revtex}

\begin{document}

\draft

\preprint{gr-qc/0207014}

\title{Removal of Closed Timelike Curves in Kerr-Newman Spacetime}

\author{Hongsu Kim\footnote{hongsu@hepth.hanyang.ac.kr}}

\address{Department of Physics\\
Hanyang University, Seoul, 133-791, KOREA}

\date{June, 2002}

\maketitle

\begin{abstract}
A simple yet systematic new algorithm to investigate the
global structure of Kerr-Newman spacetime is suggested.
Namely, the global structures of $\theta =$ const.
timelike submanifolds of Kerr-Newman metric 
are studied by introducing a new time coordinate slightly different
from the usual Boyer-Lindquist time coordinate.
In addition, it is demonstrated that the possible causality violation
thus far regarded to occur near the ring singularity via the development
of closed timelike curves there is not really an unavoidable pathology
which has plagued Kerr-Newman solution but simply a gauge (coordinate) artifact
as it disappears upon transforming from Boyer-Lindquist to the new
time coordinate. This last point appears to lend support to the fact that,
indeed, the Kerr-Newman spacetime is a legitimate solution to represent
the interior as well as the exterior regions of a rotating, charged black
hole spacetime.

\end{abstract}

\pacs{PACS numbers: 04.20.Jb, 04.20.Cv, 97.60.Lf }

\narrowtext


\begin{center}
{\rm \bf I. Introduction}
\end{center}

In the present work, we would like to suggest a new algorithm to
investigate the global structure and maximal analytic extension of
the Kerr-Newman spacetime [1,2] which is simpler yet more systematic
than the existing ones. To our knowledge, the first serious attempts
to uncover them were made long ago by Carter [5,7] and by Boyer and 
Lindquist [6]. Along the line of procedures suggested earlier by 
Finkelstein [8] and by Kruskal [9] and by Graves and Brill [4], 
Boyer and Lindquist [6] and independently Carter [5] were able
to perform a detailed study of the global structure and the maximal
analytic extension of the Kerr-Newman spacetime. Working in the
Kerr coordinates [6], which can be thought of as the
axisymmetric generalization of Eddington-Finkelstein null coordinates, 
Carter [5], however, had to confine himself to the study particularly on the
symmetry axis $\theta = 0$ (where $\theta $ denotes the polar
angle) for technical limitations. And then from the conclusion
derived from this study, he conjectured that the other $\theta
=$const. hypersurfaces would presumably have the same global
structures as that of the symmetry axis. The new algorithm we are
about to present here allows us to ``complete'' the status of
development in this direction. Namely, we shall be able to explore
the global structures of the ``$\theta =$ const." timelike
submanifolds of Kerr-Newman spacetime from the ``symmetry axis"
($\theta = 0$) all the way to the ``equatorial plane" ($\theta =
\pi/2$) one by one. Our strategy is first to scan the full, 4-dim.
Kerr-Newman manifold by slicing it into pieces of $\theta =$const.
3-dim. timelike submanifolds and next to examine the global
structure of each one of them. And to this end, we shall introduce
a ``new'' time coordinate different from the usual Boyer-Lindquist
time coordinate [6] which can be defined only on the $\theta =$const.
timelike submanifolds. Now let us be more specific. Since the
geometry of $\theta = \theta_{0}$ (with $0 < \theta_{0}\leq
\pi/2$) submanifolds is essentially 3-dim. whereas that of the
symmetry axis ($\theta_{0} = 0$) is {\it effectively} 2-dim.
(since it has a degeneracy along $\phi$ direction), things get
more involved compared to the case of the symmetry axis discussed
by Carter [5]. Therefore in order to make the analysis tractable,
we shall introduce a ``new'' time coordinate $\tilde{t}$ slightly
different from the usual Boyer-Lindquist time coordinate $t$ [6]
based on the philosophy that the global structure remains
unaffected under coordinate changes. From there one can then apply
the same methods of Finkelstein [8] and Kruskal [9] to obtain the
maximal analytic extensions of the conformal diagrams of the
$\theta = \theta_{0}$ submanifolds. Then it is straightforward to
see that the maximally extended conformal diagram of the $\theta =
\theta_{0}$ (with $0 < \theta_{0} < \pi/2$) submanifolds of
Kerr-Newman spacetime remain the same, i.e., still take the same
structure as that of the symmetry axis [5]. The extended conformal
diagram of the equatorial plane, on the other hand, takes exactly
the same structure as that of the Reissner-Nordstrom (RN) solution
[3]. As a result, it does have the ring singularity at $\Sigma =
0$ (or more precisely at $r = 0$ since $\theta_{0} = \pi/2$) and
one cannot, on the equatorial plane, extend to negative values of
$r$ while one can do so on other $\theta =\theta_{0}$
submanifolds. As we shall discuss in detail in the text shortly,
in addition to the clearer overview of the global structure of the
full Kerr-Newman spacetime that our algorithm provides, there is
another, even more crucial role played by the ``new'' time
coordinate. Recall, as was first pointed out by Carter [7], that
there is an issue of possible occurrence of closed timelike curves
and hence the possibility of causality violation near the ring
singularity if one analyzes the Kerr-Newman metric in
Boyer-Lindquist coordinates. As we shall see in a moment, this
occurrence of closed timelike curves disappears by transforming to
the ``new'' time coordinate. This, of course, implies that the
seemingly possible violation of causality near the ring
singularity of Kerr-Newman spacetime is {\it not} an unavoidable
pathology but simply a gauge (coordinate) artifact whose
appearance can be attributed to the poor choice of
(Boyer-Lindquist) coordinates. In short, this elimination of the
possibility of causality violation near the ring singularity via a
transformation to the new time coordinate appears to lead us to
believe that the Kerr-Newman solution [1,2] might be really valid
as the spacetime of a rotating charged black hole. And our study,
particularly, of the global topology of the equatorial plane
confirms the existence of the ring singularity in a more
transparent manner and supports the general belief that the
spacetime produced by physically realistic collapse of even
nonspherical bodies would be qualitatively similar to the
spherical case, i.e., the RN geometry [3].

\begin{center}
{\rm \bf II. Introduction of a ``New'' time coordinate.}
\end{center}

Now consider the stationary axisymmetric Kerr-Newman solution of
the Einstein-Maxwell equations. The Kerr-Newman metric is given in
Boyer-Lindquist coordinates as [3,6]
\begin{eqnarray}
ds^2 = &-& [{\Delta - a^2 \sin^2 \theta \over \Sigma}] dt^2 -
{2a\sin^2 \theta (r^2 + a^2 - \Delta) \over \Sigma} dt d\phi \nonumber \\
&+& [{(r^2 + a^2)^2 - \Delta  a^2 \sin^2 \theta \over \Sigma}]  \sin^2 \theta
d\phi^2 + {\Sigma \over \Delta} dr^2 + \Sigma d\theta^2
\end{eqnarray}
where $\Sigma \equiv r^2 + a^2 \cos^2 \theta$ and $\Delta \equiv
r^2 - 2Mr + a^2 + e^2$ with $M$ being the mass, $a$ being the angular
momentum per unit mass and $e$ being the total $U(1)$ charge of the hole.
We are now particularly interested in the $\theta =$const. timelike surfaces
as {\it submanifolds} of this Kerr-Newman spacetime.
Namely, consider the $\theta = \theta_{0}$ ($0 \leq \theta_{0} \leq
\pi/2$) timelike submanifolds of Kerr-Newman spacetime with the metric
\begin{eqnarray}
ds^2 = &-& [{\Delta - a^2 \sin^2 \theta_{0} \over \Sigma}] dt^2 -
{2a\sin^2 \theta_{0}(r^2 + a^2 - \Delta) \over \Sigma} dt d\phi \nonumber \\
&+& [{(r^2 + a^2)^2 - \Delta a^2 \sin^2 \theta_{0} \over \Sigma}]  \sin^2
\theta_{0}
d\phi^2 + {\Sigma \over \Delta} dr^2
\end{eqnarray}
where $\Sigma = r^2 + a^2 \cos^2 \theta_{0}$ now.
These $\theta = \theta_{0}$ surfaces have metrics which are
literally 3-dim. in structure and possess an off-diagonal component in an
intricate way. Thus in order to make the study of global structure of
$\theta = \theta_{0}$ submanifolds tractable, here we consider a coordinate
transformation which is defined {\it only} on the $\theta =$const. timelike
submanifolds. Namely consider a transformation from the Boyer-Lindquist
time coordinate ``$t$'' to a new time coordinate ``$\tilde{t}$'' given by
\begin{eqnarray}
\tilde{t} = t - (a\sin^2 \theta_{0}) \phi
\end{eqnarray}
with other spatial coordinates $(r, ~\phi)$ remaining unchanged.
Note that the new time coordinate $\tilde{t}$ is different from the old one
$t$ only for ``rotating'' case ($a \neq 0$) and even then only for
$\theta_{0} \neq 0$. In terms of this ``new'' time coordinate $\tilde{t}$,
the metric of $\theta = \theta_{0}$ submanifolds becomes
\begin{eqnarray}
ds^2 &=& - {\Delta \over \Sigma} d\tilde{t}^2 + {\Sigma \over \Delta} dr^2 +
\Sigma  \sin^2 \theta_{0} (d\phi -{a\over \Sigma}d\tilde{t})^2 \\
     &=& - N^2(r) d\tilde{t}^2 + h_{rr}(r) dr^2 +
     h_{\phi\phi}(r) [d\phi + N^{\phi}(r)d\tilde{t}]^2. \nonumber
\end{eqnarray}
Namely, the metric now takes on the structure of a remarkably
simple ADM's $(2 + 1)$ space-plus-time split form with the lapse,
shift functions and the spatial metric components being given
respectively by
\begin{eqnarray}
N^2(r) &=& {\Delta \over \Sigma},   ~~~N^{\phi}(r)=-{a \over \Sigma},
~~~N^{r}(r) = 0 \\
h_{rr}(r) &=& N^{-2}(r), ~~~h_{\phi\phi}(r) = \Sigma \sin^2 \theta_{0},
~~~h_{r\phi}(r) = h_{\phi r}(r) = 0. \nonumber
\end{eqnarray}
In terms of this new time coordinate  $\tilde{t}$, therefore, it
is now manifest that the $\theta = \theta_{0}$ submanifolds of
Kerr-Newman spacetime have the topology of $R^2 \times S^1$ (which
was rather obscured in the metric form eq.(2) given in the
original Boyer-Lindquist time coordinate $t$) and hence the new
time coordinate is better-suited for the study of global
structure. That is to say, the global structure of the timelike
2-dimensional
submanifolds ${\it T}=R^2$ would mirror that of the full $\theta =
\theta_{0}$ submanifolds since each point of  ${\it T}=R^2$ can be
thought of as representing $S^1$. We would like to add a comment
here. Of course it is true that it is the manifold itself that has
topology, not the metric. Therefore, regardless of the coordinates
and hence the metrics one chooses, they all describe the same
manifold with a single topology. However, the point we would like
to make here is that the metric given in new time coordinate
$\tilde{t}$ (eq.(4)) demonstrates more clearly that the  $\theta =
\theta_{0}$ submanifolds it describes has the topology of $R^2
\times S^1$ than the metric in Boyer-Lindquist time coordinate $t$
(eq.(2)) does. \\ Thus we now turn to the analysis of the global
structures of the $\theta = \theta_{0}$ timelike submanifolds.
Firstly, consider the
$\theta_{0} = 0$ timelike submanifolds representing the ``symmetry
axis'' of the Kerr-Newman spacetime with the metric being given by
\begin{eqnarray}
ds^2 = - ({\Delta \over  r^2 + a^2}) dt^2 + ({\Delta \over  r^2 + a^2})^{-1}
dr^2.
\end{eqnarray}
Note that this metric of the symmetry axis is effectively 2-dim.
since it is {\it degenerate} along the $\phi$ direction. And this
diagonal, 2-dim. structure of the metric of the symmetry axis
allowed a complete analysis of its global structure as had been
carried out by Carter [5]. \\ Secondly, consider the $\theta_{0} =
\pi/2$ surface which represents the ``equatorial plane'' of
Kerr-Newman spacetime. The metric of this submanifold is obtained
in the new time coordinate $\tilde{t}$ by setting
 $\theta_{0} = \pi/2$ in eq.(4)
\begin{eqnarray}
ds^2 = - N^2(r) d\tilde{t}^2 + N^{-2}(r) dr^2 +
r^2 [d\phi + N^{\phi}(r)d\tilde{t}]^2
\end{eqnarray}
with the lapse $N(r)$ and the shift $N^{\phi}(r)$ in above $(2 + 1)$-
split form being given by
\begin{eqnarray}
N^2(r) = {\Delta \over r^2} = [ 1 - {2M \over r} + {(a^2 + e^2) \over
r^2}], ~~~N^{\phi}(r) = - {a\over r^2}.
\end{eqnarray}
Finally, note that the metric of the
 $\theta = \theta_{0}$ ($0 < \theta_{0} < \pi/2$) submanifolds given
in eq.(4) are everywhere non-singular including $r =  0$ and
possess exactly the same causal structure (except for the
appearance of ergoregion) as that of the symmetry axis $(\theta =
0)$. Therefore, the maximal analytic extension of the $\theta =
\theta_{0}$ ($0 < \theta_{0} < \pi/2$) submanifolds representing
their global structure is essentially the same as that of the
symmetry axis first studied by Carter [5]. The metric of the
equatorial plane $(\theta_{0} = \pi/2)$ given in eqs.(7),(8), however,
possesses a curvature singularity at $r = 0$ as expected (since it
is the ``ring singularity'', $r = 0$, $\theta_{0} = \pi/2$)
whereas it exhibits almost the same causal structure (again except
for the presence of the ergoregion) as that of the symmetry axis.
As a result, the maximal analytic extension of the equatorial
plane is identical to that of the RN spacetime [3]. Detailed analysis
of the global structure and the maximal analytic extension will be
presented in a separate publication.

\begin{center}
{\rm \bf III. Physical meaning of the ``New'' time coordinate.}
\end{center}

Note that both in the old Boyer-Lindquist coordinates and in the
new coordinates, the $\theta =\theta_{0}$ submanifolds of
Kerr-Newmann metric are stationary and axisymmetric and hence
possess the associated time-translational and rotational Killing
fields $\{\xi^{\mu}=(\partial/\partial t)^{\mu}, \psi^{\mu}=
(\partial/\partial \phi)^{\mu}\}$ and $\{\tilde{\xi}^{\mu}=
(\partial/\partial \tilde{t})^{\mu}, \tilde{\psi}^{\mu}=
(\partial/\partial \tilde{\phi})^{\mu}\}$ respectively.
Since the azimuthal angle coordinate $\phi$ is not being transformed,
we shall henceforth write $\tilde{\phi} = \phi$ in all the
expressions given in the new coordinates $(\tilde{t}, r, \tilde{\phi})$.
Now, in order to have physical meaning of
the new coordinate system $(\tilde{t}, r, \tilde{\phi})$ for each
$\theta = \theta_{0}$ timelike submanifold given by eq.(3), we
begin by considering how the associated time-translational and
rotational Killing fields transform accordingly. Namely from :
\begin{eqnarray}
g_{\tilde{t}\tilde{t}} &=& g_{tt}, ~~~g_{\tilde{t}\tilde{\phi}} =
(a\sin^2\theta_{0})g_{tt} + g_{t\phi}, \\
g_{\tilde{\phi}\tilde{\phi}} &=& (a^2\sin^4\theta_{0})g_{tt} +
2(a\sin^2\theta_{0})g_{t\phi} + g_{\phi\phi} \nonumber
\end{eqnarray}
and $g_{tt} = \xi^{\mu}\xi_{\mu}$, $g_{t\phi} =
\xi^{\mu}\psi_{\mu}$, $g_{\phi\phi} = \psi^{\mu}\psi_{\mu}$ and
$g_{\tilde{t}\tilde{t}} = \tilde{\xi}^{\mu}\tilde{\xi}_{\mu}$,
$g_{\tilde{t}\tilde{\phi}} = \tilde{\xi}^{\mu}\tilde{\psi}_{\mu}$,
$g_{\tilde{\phi}\tilde{\phi}}
=\tilde{\psi}^{\mu}\tilde{\psi}_{\mu}$,
we can conclude
\begin{eqnarray}
\tilde{\xi}^{\mu} = \xi^{\mu}, ~~~\tilde{\psi}^{\mu} = \psi^{\mu}
+ (a\sin^2\theta_{0})\xi^{\mu}.
\end{eqnarray}
We are now ready
to interpret the physical meaning of this coordinate
transformation on each $\theta = \theta_{0}$ timelike submanifold
in terms of this transformation law for the Killing fields. In the
old Boyer-Lindquist coordinate $(t, r, \phi)$, the coordinate
``$t$'' is a timelike coordinate outside the event horizon with
unbounded range $0\leq t <\infty$ and the coordinate ``$\phi$'' is
compactified with the period of $2\pi$, i.e., $\phi \sim \phi +
2\pi n$ $(n \in Z)$ in order for the Kerr-Newman metric to be
asymptotically-flat. And this amounts to identifying points along
the ``straight'' orbits of the rotational Killing field
$(\partial/\partial \phi)^{\mu}$. Now, notice that under the
coordinate transformation given in eq.(3), the associated Killing
fields transform as in eq.(10), namely
\begin{eqnarray}
\left(\partial/\partial \tilde{\phi}\right)^{\mu} =
\left(\partial/\partial \phi\right)^{\mu} + a\sin^2\theta_{0}
\left(\partial/\partial t\right)^{\mu}, ~~~\left(\partial/\partial
\tilde{t}\right)^{\mu} = \left(\partial/\partial t\right)^{\mu}.
\end{eqnarray}
Particularly, the transformation law for the rotational Killing
field $\psi^{\mu} = (\partial/\partial \phi)^{\mu}$ indicates that
now one should identify points along the ``twisted'' orbits of
$\psi^{\mu} + (a\sin^2\theta_{0})\xi^{\mu}$. Namely, since the
compactification direction now involves that of time coordinate,
the corresponding identification of points should be
\begin{eqnarray}
(t, r, \phi) \equiv (t+2\pi n(a\sin^2\theta_{0}), r, \phi+2\pi n),
~~~n\in Z
\end{eqnarray}
which implies that, except along the poles, $t$ is periodic,
namely, there are {\it closed timelike curves}. Then upon this
change of coordinates, the metric of each $\theta = \theta_{0}$
timelike submanifold remarkably simplifies to the form given in
eq.(4) or equivalently to
\begin{eqnarray}
ds^2 = -\left[{\Delta - a^2\sin^2\theta_{0} \over \Sigma}\right]
d\tilde{t}^2 - 2a\sin^2\theta_{0}d\tilde{t}d\tilde{\phi} + \Sigma
\sin^2\theta_{0}d\tilde{\phi}^2 + {\Sigma \over \Delta}dr^2,
\end{eqnarray}
now with the identification of points $(\tilde{t}, r,
\tilde{\phi}) \equiv (\tilde{t}, r, \tilde{\phi}+2\pi n)$ in a
standard manner. The fact that one recovers standard way of point
identification after the coordinate transformation given in eq.(3)
can indeed be represented as follows : just as the old
Boyer-Lindquist time coordinate ``$t$'' is constant along the
orbits of $\psi^{\mu}$, i.e., $(\partial/\partial \phi)t = 0$, the
new coordinate ``$\tilde{t}$'' is the ``adapted'' time coordinate 
which is constant along the orbits of 
$\tilde{\psi}^{\mu} = \psi^{\mu} + (a\sin^2\theta_{0})\xi^{\mu}$, i.e.,
\begin{eqnarray}
\left(\partial/\partial \tilde{\phi}\right) \tilde{t} =
\left[\left(\partial/\partial \phi\right) + a\sin^2\theta_{0}
\left(\partial/\partial t\right)\right] (t -
a\sin^2\theta_{0}\phi) = 0.
\end{eqnarray}
As we stressed earlier, in terms of new coordinates, it becomes
much more transparent that the topology of each $\theta =
\theta_{0}$ submanifold is $R^2\times S^1$, which, although
expected, was not so clear in the old Boyer-Lindquist coordinates.
We now remark on some interesting features that emerge when we
describe the $\theta = \theta_{0}$ submanifolds of the Kerr-Newman
black hole spacetime in terms of this new coordinates aside from
the advantage in studying their global structure. \\ (1) {\it
Gauge (Coordinate) dependence of the occurrence of closed timelike
curves}. \\ Indeed, this is the point of central importance we
would like to make in the present work concerning the crucial role
played by this new coordinate system. First, notice that the
transformation from the Boyer-Lindquist time coordinate ``$t$'' to
the new time coordinate ``$\tilde{t}$'' introduced in the present
work involves the identification of points $t \sim t + 2\pi
n(a\sin^2\theta_{0})$ in addition to the usual $\phi \sim \phi +
2\pi n$. As was pointed out, this implies that this particular
transformation involves, along the way, the {\it compactification
of time coordinate} or equivalently, the {\it emergence of closed
timelike curves} away from the poles at $\theta_{0} = 0, \pi$.
Once the transformation to the new time coordinate is completed,
however, the $\theta = \theta_{0}$ submanifolds of the Kerr-Newman
spacetime admits metrics free of the closed timelike curves since
now $\tilde{t}$ possesses usual, semi-infinite, unbounded range,
$0\leq \tilde{t} <\infty$. Namely, this coordinate transformation
partly involves an implicit process in which the closed timelike
curves are ``gauged away''. One might suspect that the
compactification of old time coordinate involved in this
coordinate transformation is not relevant and hence physically
unacceptable. It is, however, well-known that closed timelike
curves indeed occur in the interior region of Kerr-Newman black
hole spacetime represented in Boyer-Lindquist coordinates and
hence the explicit emergence of the closed timelike curves during
the course of this coordinate transformation process presumably
could reflect this fact. Thus in the following, we shall
demonstrate that this is indeed the case and hence in the new time
coordinate $\tilde{t}$, the closed timelike curves are completely
gauged away. To this end, we begin by recalling the argument first
given by Carter [7] exhibiting the possible occurrence of causality
violation in some interior region of Kerr-Newman spacetime. \\
Consider the norm of the Killing field $\psi^{\mu} =
(\partial/\partial \phi)^{\mu}$ evaluated on a $\theta =
\theta_{0}$ $(0\leq \theta_{0} \leq \pi/2)$ submanifold with
metric given in Boyer-Lindquist coordinates ;
\begin{eqnarray}
\psi^{\mu}\psi_{\mu} = g_{\phi\phi} = \left[r^2 + a^2 +
{(2Mr-e^2)a^2\sin^2\theta_{0}\over \Sigma}\right]\sin^2\theta_{0}.
\end{eqnarray}
For ``negative'' values of $r$ of sufficiently small magnitude and
for $\theta_{0}$ sufficiently close to $\pi/2$,
$\psi^{\mu}\psi_{\mu}<0$, i.e., $\psi^{\mu} = (\partial/\partial
\phi)^{\mu}$ can go ``timelike'' near the ring singularity at
$r=0$, $\theta_{0}=\pi/2$. However, since the orbit of
$\psi^{\mu}$ must be, as stated, closed with period of $2\pi$ in
order for the Kerr-Newman spacetime to be asymptotically-flat as
$r\rightarrow \infty$, this timelike-going behavior of
$\psi^{\mu}$ near the ring singularity indicates the possible
occurrence of {\it closed timelike curves} and hence the possible
violation of causality there. On the other hand, consider this
time the norm of $\tilde{\psi}^{\mu} = (\partial/\partial
\tilde{\phi})^{\mu}$ evaluated again on the same $\theta =
\theta_{0}$ hypersurface with the metric given in new coordinates
;
\begin{eqnarray}
\tilde{\psi}^{\mu}\tilde{\psi}_{\mu} =
g_{\tilde{\phi}\tilde{\phi}} = (r^2 + a^2\cos^2\theta_{0})
\sin^2\theta_{0} > 0.
\end{eqnarray}
Namely, it is always positive-definite meaning that
$\tilde{\psi}^{\mu}$ is everywhere spacelike and can never go
timelike even near the ring singularity. Therefore, if one works
in this new coordinates $(\tilde{t}, r, \tilde{\phi})$, no
possibility of causality violation is encountered anywhere he/she
goes in the Kerr-Newman spacetime ! To conclude, the possible
timelike-going behavior of the rotational Killing field and hence
the seemingly possible causality violation near the ring
singularity turns out to be just a {\it gauge artifact}, i.e., it
can be attributed to the poor choice of (Boyer-Lindquist)
coordinates and can be eliminated by transforming to the new
coordinates introduced in the present work in much the same way as
the coordinate singularity at the event horizon of the Kerr-Newman
metric can be gauged away by transforming to null coordinates such
as Kerr coordinates or Kruskal-Szekers coordinates. As is well
understood, since the gauge, or equivalently, coordinate
transformation from one to another in gravity amounts to the shift
of observer's ``state of motion'' from one to another in curved
spacetime, this apparent gauge dependence of the occurrence of
closed timelike curves indicates that they, like the appearance of
event horizon, may be present or absent depending on the
observer's state of motion. In fact, there is a simple observation
that might provide a clue to the nature of relative motion between
the two observers representing the two coordinate systems at each
$\theta =\theta_{0}$ hypersurface. To this end, we begin with the
computation of the angular velocity of 
zero-angular-momentum-observer (ZAMO) [10] in each of the two coordinates.
Generally, the orthonormal tetrad is a set of four mutually
orthogonal unit vectors at each point in a given spacetime that give
the directions of the four axes of locally Minkowskian coordinate 
system. And the ZAMO frame (or the locally-non-rotating-frame (LNRF))
is one such orthonormal tetrad that singles out since an observer at
rest in it has zero angular momentum with respect to the rotating  
Kerr-Newman black hole. Also, ZAMO is a fiducial observer (FIDO) moving
with the 4-velocity which is orthogonal to spacelike hypersurfaces
$t =$const. Namely, the ZAMO geodesic is defined by $ds^2 = - N^2dt^2$ or
equivalently $\{dr=0, d\theta=0, (d\phi + N^{\phi}dt)=0\}$ for 
Kerr-Newman spacetime given in Boyer-Lindquist coordinates and as a
result, ZAMO angular velocity is given by $\Omega = d\phi/dt = -N^{\phi} = 
-{g_{t\phi}/g_{\phi\phi}}$. Thus the ZAMO angular velocity when evaluated
in each of the two coordinates is given by
\begin{eqnarray}
\Omega = -{g_{t\phi}\over g_{\phi\phi}} = {a(r^2+a^2-\Delta)\over {(r^2+a^2)^2-\Delta
a^2\sin^2\theta_{0}}}, 
~~~\tilde{\Omega} = -{g_{\tilde{t}\tilde{\phi}}\over g_{\tilde{\phi}\tilde{\phi}}}
= {a\over {r^2+a^2\cos^2\theta_{0}}}.
\end{eqnarray}
Next, the angular velocity of the horizon (and hence the black hole) is nothing
but that of ZAMO at $r = r_{+}$ and again in each of the two coordinates, it is
\begin{eqnarray}
\Omega_{H} = -{g_{t\phi}\over g_{\phi\phi}}\vert_{r_{+}} = {a\over
r^2_{+} + a^2}, ~~~\tilde{\Omega}_{H} =
-{g_{\tilde{t}\tilde{\phi}}\over
g_{\tilde{\phi}\tilde{\phi}}}\vert_{r_{+}} = {a\over r^2_{+} +
a^2\cos^2\theta_{0}}
\end{eqnarray}
where $r_{+} = M+(M^2-a^2-e^2)^{1/2}$ denotes the location of the
event horizon which turns out to be precisely the same (we shall
come back to this issue shortly) in the two coordinates. As we
shall see in a moment, the static limit, i.e., the outer boundary
of the ergoregion, turns out to occur exactly at the same point
$r_{s} = M + (M^2-a^2\cos^2\theta_{0}-e^2)^{1/2}$ in the two
coordinates as well. It is now straightforward to see that 
$\tilde{\Omega} > \Omega$ and $\tilde{\Omega}_{H} > \Omega_{H}$
for $r\geq r_{+}$ and $\theta_{0}\neq 0, \pi$.
Thus if we combine these observations, i.e., the
invariance of the causal structure and 
$\tilde{\Omega} > \Omega$ and hence
$\tilde{\Omega}_{H}>\Omega_{H}$ (for $\theta_{0}\neq 0, \pi$), we
can conclude that the two observers associated with the two
coordinate systems have no relative radial motions but just
relative azimuthal angular motions and the new coordinates (and
the observer in it) appears to rotate in opposite direction around
the hole's rotation axis with respect to the Boyer-Lindquist
coordinates (and the observer in it) at the angular velocity that
changes with $\theta_{0}$ such that it gets maximized at the
equatorial plane $\theta_{0}=\pi/2$ while becomes zero on the
poles $\theta_{0}=0, \pi$. To summarize, the closed timelike
curves and the event horizon are on {\it equal footing} in that
they have {\it apparent gauge dependence} and hence should be
distinguished from the unavoidable curvature singularity, i.e.,
the ring singularity at $r=0, \theta_{0}=\pi/2$ that clearly is
the drawback of the Kerr-Newman solution as a classical solution
to the Einstein-Maxwell equations. Indeed, due to the occurrence
of closed timelike curves and hence the emergence of causality
violation near the ring singularity, the Kerr-Newman solution has
been regarded as not being able to represent the interior region
of charged, rotating black holes thus far. In this regard, the
demonstration of the possible elimination of closed timelike
curves near the ring singularity (and elsewhere) via the
transformation to the new coordinates discovered in the present
work appears to lend support to the viability of the Kerr-Newman
solution as a legitimate, physical metric being able to represent
both the exterior and interior of charged, rotating black hole
spacetime. \\ (2) {\it Invariance of the causal structure} \\
Note that it is the linear combinations
\begin{eqnarray}
\chi^{\mu} = \xi^{\mu} + \Omega_{H} \psi^{\mu},
~~~\tilde{\chi}^{\mu} = \tilde{\xi}^{\mu} + \tilde{\Omega}_{H}
\tilde{\psi}^{\mu}
\end{eqnarray}
constructed out of the time-translational and
rotational Killing fields $\{\xi^{\mu}, \psi^{\mu}\}$ and
$\{\tilde{\xi}^{\mu}, \tilde{\psi}^{\mu}\}$ in Boyer-Lindquist and
new coordinates respectively
which are tangential to the null geodesic generator
of the horizon. And $\Omega_{H}$ and $\tilde{\Omega}_{H}$ above
are as given earlier in eq.(18). Now, the event horizon is a
Killing horizon and generally speaking, Killing horizons occur at
points where the KIlling field $\chi^{\mu}$ or
$\tilde{\chi}^{\mu}$ above becomes null. As is well-known,
$\chi^{\mu}\chi_{\mu}=0$ occurs for $\Delta = r^2-2Mr+a^2+e^2=0$,
namely at $r_{\pm} = M\pm (M^2-a^2-e^2)^{1/2}$. Interestingly, but
as expected to some extent, even in the new coordinates,
$\tilde{\chi}^{\mu}\tilde{\chi}_{\mu}=0$ occurs again for $\Delta
= 0$. Next, we turn to the static limit, the outer boundary of the
ergoregion. It occurs at the point where the time-translational
Killing field $\xi^{\mu}$ or $\tilde{\xi}^{\mu}$ becomes null.
Thus again, it is amusing to note both in the old Boyer-Lindquist
and in the new coordinates, the static limit occurs exactly at the
same point $r_{s} = M + (M^2-a^2\cos^2\theta_{0}-e^2)^{1/2}$ since
$\xi^{\mu}\xi_{\mu} = \tilde{\xi}^{\mu}\tilde{\xi}_{\mu} = g_{tt}
= -\left[{\Delta-a^2\sin^2\theta_{0} \over \Sigma}\right] = 0$
occurs for $(\Delta-a^2\sin^2\theta_{0}) = 0$. To summarize, the
Killing horizons and the static limit occur respectively at
precisely the same locations and hence the causal structure of the
Kerr-Newman black hole spacetime remains the same regardless of
the coordinate systems (old or new) used to represent the metrics
of $\theta = \theta_{0}$ submanifolds. And of course, it can be
attributed to the nature of the coordinate transformation in which
the radial coordinate ``$r$'' remains unaffected.

\begin{center}
{\rm \bf IV. Concluding Remarks}
\end{center}

We now conclude with some comments worth mentioning. First, we
would like to point out the complementary roles played by the two
alternative time coordinates $t$ and $\tilde{t}$. The
Boyer-Lindquist time coordinate $t$ is the usual Killing time
coordinate with which one can obtain the rotating hole's
characteristics such as the angular velocity of the event horizon
or the surface gravity as measured by an outside observer who is
``static" with respect to, say, a distant star. The ``new" time
coordinate $\tilde{t}$, on the other hand, is a kind of an unusual
one in that it can be identified with the time coordinate of a
non-static frame which rotates around the axis of the Kerr-Newman
black hole in opposite direction to that of the hole (of course
outside the ergoregion) with an angular velocity that increases
with the polar angle. This new time coordinate, however, is
particularly advantageous in exploring the global stucture of the
$\theta =$ const. submanifolds of Kerr-Newman black hole since it
allows one to transform to Kruskal-type coordinates and hence
eventually allows one to draw the Carter-Penrose diagrams much
more easily than the case when one employs the usual
Boyer-Lindquist time coordinate. Yet, however, the most remarkable
point to be emphasized concerning the important roles played by
the new time coordinate $\tilde{t}$ is that in terms of which the
seemingly possible causality violation near the ring singularity
can be eliminated as we pointed out earlier. Namely, consider the
norm of the rotational Killing field
$\tilde{\psi}^{\mu}=(\partial/\partial \tilde{\phi})^{\mu}$
evaluated on a $\theta = \theta_{0}$ ($0\leq \theta_{0}\leq
\pi/2$) timelike submanifold. Unlike the case when one employs the
usual Boyer-Lindquist time coordinate $t$, if one employs the new
time coordinate $\tilde{t}$, then
$\tilde{\psi}^{\mu}\tilde{\psi}_{\mu} =
g_{\tilde{\phi}\tilde{\phi}} > 0$, i.e., $\tilde{\psi}^{\mu}$ is
everywhere spacelike and can never go timelike even near the ring
singularity. This observation implies that the possible occurrence
of closed timelike curves and hence the possibility of causality
violation in the vicinity of ring singularity is just a gauge
artifact as it disappears upon a gauge (coordinate)
transformation. This last point appears to be a big discovery and
hence the introduction of the ``new'' time coordinate $\tilde{t}$
leads us to regard the Kerr-Newman spacetime as a valid solution
to describe both the exterior and the interior regions of a
rotating charged black hole with ever more credibility.

\begin{center}
{\rm \bf Acknowledgement}
\end{center}

This work was supported in part by BK21 project in physics department at
Hanyang university and by grant No. R01-1999-00020 from the Korea
Science and Engineering Foundation.

\vspace{2cm}

\begin{center}
{\rm \bf \large References}
\end{center}

\begin{description}

\item {[1]} R. P. Kerr, Phys. Rev. Lett. {\bf 11}, 237 (1963) ;
            R. P. Kerr and A. Schild, Am. Math. Soc. Symposium, New York, 1964.
\item {[2]} E. T. Newman and A. I. Janis, J. Math. Phys. {\bf 6}, 915 (1965) ;
E. T. Newman, E. Couch, R. Chinnapared, A Exton, A. Prakash, and R. Torrence,
J. Math. Phys. {\bf 6}, 918 (1965).
\item {[3]} S. W. Hawking and G. F. R. Ellis, {\it The Large Scale Structure of Space-Time}
            (Cambridge University Press, 1973).
\item {[4]} J. C. Graves and D. R. Brill, Phys. Rev. {\bf 120}, 1507 (1960).
\item {[5]} B. Carter, Phys. Rev. {\bf 141}, 1242 (1966).
\item {[6]} R. H. Boyer and R. W. Lindquist, J. Math. Phys. {\bf 8}, 265 (1967).
\item {[7]} B. Carter, Phys. Rev. {\bf 174}, 1559 (1968).
\item {[8]} D. Finkelstein, Phys. Rev. {\bf 110}, 965 (1958).
\item {[9]} M. D. Kruskal, Phys. Rev. {\bf 119}, 1743 (1960).
\item {[10]} J. M. Bardeen, Astrophys. J. {\bf 162}, 71 (1970) ; J. M. Bardeen,
            W. H. Press, and S. A. Teukolsky, {\it ibid.} {\bf 178}, 347 (1972).

\end{description}

\end{document}